\documentclass[conference]{IEEEtran}
\IEEEoverridecommandlockouts
\usepackage{cite}
\usepackage{amsmath,amssymb,amsfonts}
\usepackage{algorithmic}
\usepackage{graphicx}
\usepackage{textcomp}
\usepackage{amsmath}
\usepackage{xcolor}
\usepackage{multirow}
\usepackage{booktabs}
\usepackage{caption}
\usepackage{tikz}
\usepackage{url}
\usepackage{comment}
\usepackage{amssymb}
\usepackage{comment}

\def\BibTeX{{\rm B\kern-.05em{\sc i\kern-.025em b}\kern-.08em
    T\kern-.1667em\lower.7ex\hbox{E}\kern-.125emX}}
\begin{document}

\title{CATSE: A Context-Aware Framework for Causal Target Sound Extraction

\thanks{Work performed while Shrishail Baligar was a Co-op at Bose Corporation.}
}
\author{
  \IEEEauthorblockN{Shrishail Baligar\IEEEauthorrefmark{1,2}, Mikolaj Kegler\IEEEauthorrefmark{2}, Bryce Irvin\IEEEauthorrefmark{2}, Marko Stamenovic\IEEEauthorrefmark{2}, Shawn Newsam\IEEEauthorrefmark{1}} \\
  \IEEEauthorblockA{
    \IEEEauthorrefmark{2}Bose Corporation, USA\\
    \IEEEauthorrefmark{1}Electrical Engineering and Computer Science, University of California, Merced
  }
}

\maketitle
\begin{abstract}
Target Sound Extraction (TSE) focuses on the problem of separating sources of interest, indicated by a user's cue, from the input mixture. Most existing solutions operate in an offline fashion and are not suited to the low-latency causal processing constraints imposed by applications in live-streamed content such as augmented hearing. We introduce a family of context-aware low-latency causal TSE models suitable for real-time processing. First, we explore the utility of context by providing the TSE model with oracle information about what sound classes make up the input mixture, where the objective of the model is to extract one or more sources of interest indicated by the user. Since the practical applications of oracle models are limited due to their assumptions, we introduce a composite multi-task training objective involving separation and classification losses. Our evaluation involving single- and multi-source extraction shows the benefit of using context information in the model either by means of providing full context or via the proposed multi-task training loss without the need for full context information. Specifically, we show that our proposed model outperforms size- and latency-matched Waveformer, a state-of-the-art model for real-time TSE. 
\end{abstract}

\begin{IEEEkeywords}
Target sound extraction, context awareness, source separation, deep neural networks 
\end{IEEEkeywords}

\section{Introduction}

The objective of the target sound extraction (TSE) task is to separate one or more sources of interest from the input mixture~\cite{ochiai2020listen,delcroix2022soundbeam, baligar2022cossd}. The specific sounds to be extracted are specified by a query (often referred as a cue or a hint). This query can take the form of an audio clip \cite{baligar2022cossd}, text \cite{kilgour2022text}, images \cite{gao2019co}, one- or multi-hot vectors \cite{delcroix2022soundbeam}, 
etc. 

Most existing TSE methods are non-causal and thus mainly suited for processing audio segments offline for applications in music production and audiovisual media post-production \cite{slizovskaia2021conditioned}. To enable the use of TSE methods in real-time applications involving streaming audio, the models must be causal and incur minimal latency. For applications such as augmented hearing in wearable devices like hearing aids and augmented reality headsets, the total latency is typically required to be below 10 ms~\cite{yang1997echo}. Recently-proposed Waveformer~\cite{veluri2023real} is one of the unique methods suitable for real-time TSE applications for general sound classes, and represents the current state-of-the-art performance in its class.

Baligar \& Newsam~\cite{baligar2022cossd} showed that TSE tasks can be compartmentalized into two sub-tasks of target source detection and separation. Models optimized jointly on the two tasks yielded a significant performance improvement, as compared to the typical TSE baseline using only the separation objective. However, their proposed method addresses the problem of offline, non-casual TSE, and thus cannot be directly applied in a real-time causal system. 
We hypothesize that providing \textit{context-awareness}, achieved through analyzing the composition of the input mixture, can benefit causal, low-latency TSE models.

In this paper, we present 
causal, low-latency Context-Aware TSE models (CATSE) based on temporal convolutional networks (TCN) \cite{lea2016temporal}. To assess the utility of context awareness in real-time TSE models, we first provide the model with information about the ground-truth (oracle) set of classes comprising the input mixture. Such \textit{explicit} CATSE (eCATSE) provides an estimate of the upper-bound of performance, assuming the composition of the input mixture is known. While the assumption about the known set of sound classes comprising the input mixture is useful for analyzing model architectures, it is not a practical assumption for most real-time TSE models.

\begin{figure*}
  \centering
  \includegraphics[width=0.95\linewidth]{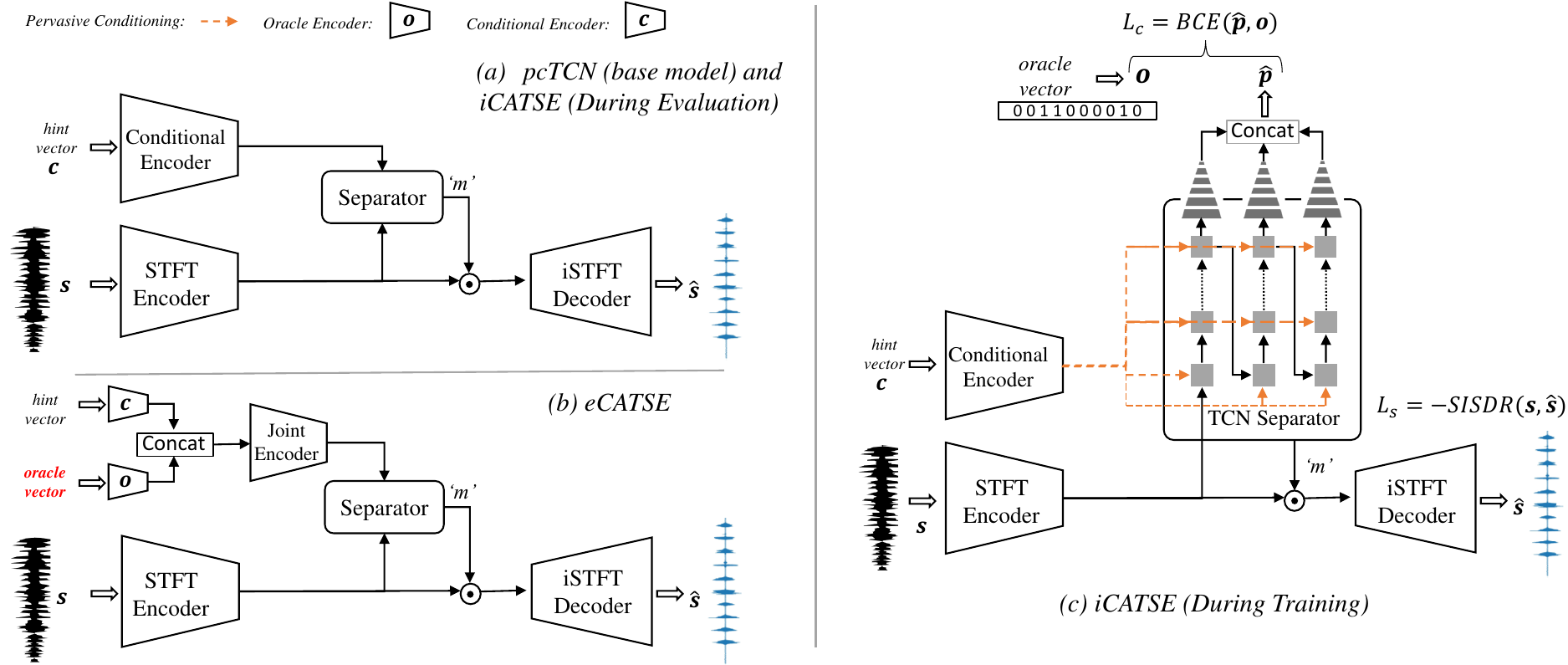}
  \caption{Overview of the proposed causal Context-Aware TSE models. (a) pcTCN applies conditioning to the separator. (b) eCATSE integrates oracle context in addition to the hint. (c) iCATSE achieves context awareness during multi-task training through classification heads, which are not used for inference.
  }
  \label{fig:main_diagram}
\end{figure*}

To overcome this limitation we propose the \textit{implicit} CATSE model (iCATSE). In particular, we aim to incorporate the identification of the sound classes comprising the input mixture as an additional training objective for the model, in addition to the TSE separation loss. We accomplish this by aggregating latent activations of subsequent TCN blocks to perform multi-class classification of the sound categories present in the input. This extension of the model is used only during training and the classification heads are discarded during inference, preserving causality at inference time.

Using objective metrics, we demonstrate that incorporating context awareness into the TSE models, both explicitly (eCATSE) and implicitly (iCATSE) facilitates the multi-class separation performance. While expected in the case of eCATSE due to the use of oracle context cues, the iCATSE variant also yielded improvements due to the multi-objective training, which supports our hypothesis. In fact, both of the proposed context models 
outperform the Waveformer model for real-time TSE.

\section{Related Work}

\noindent\textbf{Non-causal Target Sound Extraction:} Separating target sounds from a mixture by constraining the source separators has been explored in the past. Delcroix et al. propose the Soundbeam model~\cite{delcroix2022soundbeam}, which uses both class labels and enrollment/audio clips for robust conditioning. Soundbeam also accounts for inactive classes, new classes, and multi-target extraction from a mixture. 
Gfeller et al.~\cite{gfeller2021one} propose \textit{SoundFilter}, a one-shot learning based model to separate sound classes that the model has never seen before using a U-Net. Kong et al \cite{kong2020source} propose using a sound event detector to inform their separation model. This model also claims to be a USS (Universal Source Separation)~\cite{kavalerov2019universal} system that can separate hundreds of sound classes
using a single model. The mentioned methods show flexibility in addressing the TSE problem, however, they all operate in the non-causal, offline setting which is not suitable for low-latency applications. 

\noindent\textbf{Low-latency, causal Target Sound Extraction:}
While causal TSE for extraction of arbitrary sounds is relatively under-explored, causal, low-latency speech extraction problem has been investigated for decades~\cite{zheng2023sixty}, in large part due to its relevance to hearing assistance \cite{fedorov2020tinylstms}. Liu \& Wang~\cite{liu2020causal} propose a Dense-U-Net based causal model for monaural talker-independent speaker separation. Their proposed speaker-number-independent training allows for realistic scenarios when the speaker number is not given beforehand. Further, Li, et al. propose low-latency real-time continuous speech separation with their Skipping Memory (SkiM) model~\cite{li2023predictive} and follow-up with a contrastive predictive coding (CPC) method~\cite{li2023predictive} applied to the previous SkiM model. Last, speech enhancement in the causal setting is investigated by Liu, et al.~\cite{liu2022separation} who propose SI-Net which performs a two-stage speech enhancement and models speech and interference information simultaneously. 

Some of these speech separation methods address separating arbitrary talkers in the mixture. Nevertheless, there does not exist any causal real-time method other than Waveformer, to our knowledge, that allows the flexibility of separating user-conditioned target of interest from an input mixture or arbitrary sounds. This work uses learned convolutional and transpose convolutional layers as encoders for processing incoming waveforms, and decoders for reconstructing waveforms, respectively. For audio separation, it uses a stack of dilated causal convolution (DCC) layers as the encoder and a transformer network as the decoder. The model performs label integration after the first transformer block in the separator's decoder. Our work proposes an alternative approach based on the TCN and utilizing context information from the input mixture. Through a comparative analysis with the Waveformer we highlight the strength of our proposed approach, point out remaining limitations, and outline avenues for future research.

\begin{figure}
    \centering
   \includegraphics[width=0.95\linewidth]{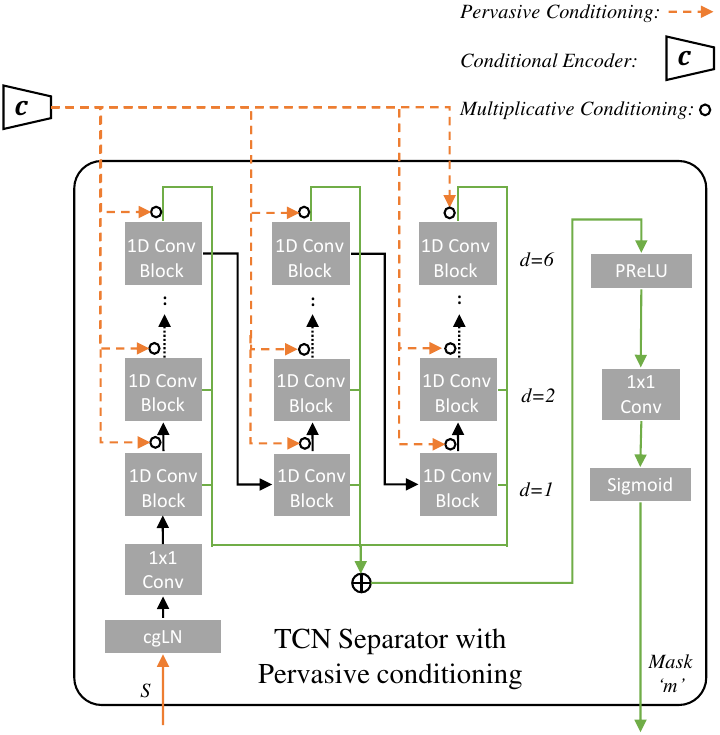}
   \caption{Our proposed TCN-based separator performs conditioning at every Conv1D layer of the three TCNs, for a total of 18 locations. This pervasive conditioning is performed in all three proposed models: pcTCN, eCATSE, and iCATSE.}
   \label{fig:pervasive}
 \end{figure}
 
\section{Method}

\subsection{Pervasively-conditioned TCN}

We introduce a pervasively-conditioned TCN (pcTCN, Fig.~\ref{fig:main_diagram}a) as a base causal TSE model that we later extend to integrate context-awareness. Our proposed framework utilizes spectral masking to extract sources of interest. A separator module receives real and imaginary components of the short-time Fourier transform (STFT) stacked along the channel axis. The resulting mask is then applied to the input and the inverse STFT is used to reconstruct the separated waveform.
We use the Asteroid \cite{Pariente2020Asteroid} implementation of STFT filterbanks with a filter size of 256, a kernel size of 128, and stride of 64. With the above-outlined setup, the algorithmic latency of our method is equivalent to the synthesis window size \cite{wang2022stft}. With the sampling rate of 16 kHz and kernel size of 128 this equates to 8 ms.

The separator is comprised of three causal TCN stacks with a depth of six 1D Conv. blocks each (Fig.~\ref{fig:pervasive}). Each individual block includes an initial 1x1 convolution expanding input to hidden channels, followed by a PReLU activation, Cumulative global layer normalization (cgLN)~\cite{luo2019conv}, and a depth-wise causal 1D convolution. It concludes with another PReLU and cgLN, then branches into a residual connection and a skip connection, both via 1x1 convolutions. All convolutions in the network are left-padded and use cgLN as a layer normalization to assure causality. The output of the model (mask \textit{m}) is obtained by summing skip connection from all the 1D conv. blocks across all the TCN stacks, followed by PReLU nonlinearity, 1x1 convolution and the sigmoid activation.

The stacked real and imaginary STFT features ${S}$ are fed into the separator alongside a conditional embedding ${c}$, indicating the target source to be extracted. The conditional encoder contains a fully connected layer with ReLU activation, which takes input as one-hot hint information and outputs an embedding vector of length 128. The conditioning information is multiplied by the outputs of each 1-D convolutional block across all TCNs, resulting in pervasive conditioning as depicted in Fig.~\ref{fig:pervasive}. There are a total of 18 conditioning points (three TCN stacks, each containing six blocks). 

The primary training objective is to minimize the negative scale-invariant signal-to-distortion ratio (SI-SDR)~\cite{le2019sdr}:
\begin{equation}
L_{s} = -10 \cdot \log_{10} \left( \frac{\| \alpha s \|^{2}}{\| \alpha s - \hat{s} \|^{2}} \right)
\end{equation}

\noindent where, $\hat{s}$ is the separated audio signal predicted by the model, $s$ is the ground truth audio signal and  \(\alpha\) is the scaling factor that aligns \(\hat{s}\) with \(s\) in terms of amplitude.

\subsection{eCATSE: Explicit Context-Aware TSE}

We provide our pcTCN model with oracle knowledge of the sounds present in a mixture to see how this enhances the separator's ability to discern and isolate target sources accurately. As shown in Fig.\ref{fig:main_diagram}(b), eCATSE has an additional encoder that takes the oracle context ${o}$ as input, which is a multi-hot vector and creates an embedding representing the classes present in the mixture. The oracle encoder is identical to the conditional encoder and also creates an embedding vector of length 128. This oracle context is concatenated with the conditional embedding ${c}$, which is later fed to a composite conditional encoder whose output is ${oc}$, an embedding meant to provide the TCN separator with a broader understanding of the sound scene. The training objective for eCATSE remains unchanged from pcTCN.

We note here that eCATSE's dependence on oracle information comes with limited practical applicability but remains relevant for tasks where all sound sources are predetermined, for example, music source separation. In this case, we may know exactly what instruments are present in a given arrangement or performance, and eCATSE is well-suited to exploit that knowledge. eCATSE also serves as an informative upper-bound for performance improvement that results from having access to this knowledge. The expected improvement is predicated on the idea that by alleviating the ambiguity of which individual sound sources make up the mixture, the model can focus entirely on perfecting the sound extraction.

\begingroup
\renewcommand{\arraystretch}{1.20} 
\begin{table*}[]
\centering
\caption{Multi-target sound extraction performance of the proposed models and baselines. SI-SNRi indicates SI-SNR improvement compared to the unprocessed, noisy clip. 
Results for the Waveformer models were obtained from the official implementation by Veluri, et al.~\cite{veluri2023real}$^{1}$. 
$^\dagger$ - model using oracle context information.
}
\label{table_performance}
\label{tab:my-table}
\resizebox{0.95\textwidth}{!}{%
\begin{tabular}{lccccccc}
\toprule
\multicolumn{1}{l}{\multirow{2}{*}{\textbf{Model}}} &
  \multirow{2}{*}{\textbf{\#Params}} &
  \multirow{2}{*}{\textbf{Latency}} &
  \multicolumn{1}{c|}{\multirow{2}{*}{\textbf{Sampling rate}}} &
  \multicolumn{4}{c}{\textbf{SI-SNRi (dB) / SNR (dB)}} \\
\multicolumn{1}{c}{} &
   &
   &
  \multicolumn{1}{c|}{} &
  \textbf{1 target} &
  \textbf{2 targets} &
  \textbf{3 targets} &
  \textbf{Avg.} \\ \hline
eCATSE$^\dagger$ (ours) &
  3.54M &
  8 ms &
  \multicolumn{1}{c|}{16 kHz} &
  12.45 / 15.09 &
  7.93 / 9.02 &
  5.62 / 6.07 &
  8.66 / 10.06 \\ \hline
iCATSE (ours) &
  3.52M &
  8 ms &
  \multicolumn{1}{c|}{16 kHz} &
  \textbf{10.07 / 13.02} &
  \textbf{4.97 / 6.44} &
  \textbf{2.26 / 2.84} &
  \textbf{5.77 / 7.43} \\
pcTCN (ours) &
  3.52M &
  8 ms &
  \multicolumn{1}{c|}{16 kHz} &
  9.79 / 12.94 &
  4.69 / 6.26 &
  1.98 / 2.60 &
  5.49 / 7.27 \\
Waveformer~\cite{veluri2023real}  &
  3.62M &
  10 ms &
  \multicolumn{1}{c|}{16 kHz} &
  9.39 / 12.77 &
  4.65 / 6.33 &
  1.30 / 2.05 &
  5.11 / 7.05 \\ \hline\hline
Waveformer~\cite{veluri2023real} &
  3.88M &
  10 ms &
  \multicolumn{1}{c|}{44.1 kHz} &
  9.29 / 12.76 &
  4.92 / 6.51 &
  1.35 / 2.04 &
  5.19 / 7.10 \\
  \bottomrule
\end{tabular}%
}
\end{table*}
\endgroup

\subsection{iCATSE: Implicit Context-Aware TSE}

In order to make Context-Aware TSE more broadly applicable to scenarios without known sources, we investigate multi-task training for implicit context awareness. The first task of iCATSE is separation;  the second task is multi-class classification meant to identify all classes present in a given mixture.
The goal of the approach is to encourage the model to analyze the contents of the input mixture and thus implicilty develop context awareness. This differs from eCATSE which uses oracle context as conditioning information in the TCN.

During training, iCATSE has three convolutional heads on top of each of the three TCN stacks, as illustrated in \ref{fig:main_diagram}(c). The three convolutional heads consists of four blocks, each consisting of a 1D convolution followed by LeakyReLU activation and Max Pooling, the outputs of which are concatenated and flattened into an embedding. This embedding is fed to a multi-class classifier containing two fully-connected layers, to predict the sound classes present in the mixture $s$. After training, these heads are discarded, maintaining a model size equal to our core model, pcTCN. 

To train the iCATSE model, we combine the separation loss with a classification loss $L_{c}$. The classification task uses binary cross-entropy (BCE) as the loss function $L_{c}$, aimed at identifying the presence or absence of specific sound classes in the mixture. 
The BCE loss for a single instance can be expressed as:

\begin{equation}
L_{c} = -\left[ o \log(\hat{p}) + (1 - o) \log(1 - \hat{p}) \right]
\end{equation}

\noindent where, $\hat{p}$ is the prediction from the classification head, and ${o}$ is oracle ground truth.

The combined loss function, $L_{sc}$, for iCATSE is thus represented by:
\begin{equation}
L_{sc} = L_{s} + \alpha \cdot L_{c}
\end{equation}

\noindent where $L_{s}$ is the SI-SDR loss for the separation task, $L_{c}$ is the BCE loss for the classification task, and $\alpha$ is a weighting factor for the two objectives. Empirical results indicate that setting $\alpha = 0.5$ gives the best separation, while still optimizing for both separation quality and classification accuracy.

\section{Results}

\subsection{Experimental setup}

We adopt the dataset and evaluation setup for single and multi-target sound extraction from the official implementation of Waveformer\footnote{\url{https://github.com/vb000/Waveformer}} to ensure that any performance gains are attributed to our method rather than dataset variations. Specifically, we use the KaggleFSD2018 \cite{fonseca2018general} dataset for foreground events and the TAU Urban Acoustic Scenes 2019 \cite{mesaros2018multi} dataset for background sounds to generate complex audio mixtures. The FSDKaggle2018 dataset contains 41 sound event classes from the AudioSet ontology \cite{gemmeke2017audio}, and is combined with the TAU Urban Acoustic Scenes dataset via Scaper \cite{salamon2017scaper}, creating a dataset of 50k training, 10k testing, and 5k validation samples. For each dataset, we are using the recommended train/validation/test split. 

Sound mixtures are created by randomly sampling foreground sounds from 3-5 unique sound classes without replacement. These foreground sounds are further cropped into 3-5 second samples and overlayed onto a 6-second long randomly selected background track. The signal-to-noise ratio (SNR) of foreground sounds is set between 15 and 25 dB relative to the background. Target sounds in each mixture are selected with a methodology consistent with Waveformer's.

We conduct all of our experiments using audio samples at 16 kHz instead of 44.1 kHz used in the original Waveformer's implementation. To ensure fair comparison of the models, we use a Waveformer model trained and evaluated using a 16kHz dataset. Nonetheless, we report results for the original 44.1kHz Waveformer for completeness. 

\begingroup
\renewcommand{\arraystretch}{1.15} 
\begin{table}[]
\centering
\caption{Single-target sound extraction performance of the proposed causal models and baseline. All the models in the table have the same configurations as in Table~\ref{table_performance} but has been re-trained to perform the single-target TSE task.}
\label{table_comparison}
\resizebox{0.85\columnwidth}{!}{%
\begin{tabular}{lcc}
\toprule
\textbf{Model}     & \textbf{SI-SNRi (dB)} & \textbf{SNR (dB)} \\ \hline
eCATSE$^\dagger$ (ours)      & 11.22                 & 13.96             \\ \hline
iCATSE (ours)      & 9.53                  & \textbf{13.28}    \\
pcTCN (ours)       & \textbf{9.82}         & 13.15             \\
Waveformer (16k)   & 9.22                  & 13.05             \\ \hline\hline
Waveformer (44.1k) & 9.43                  & 12.86  \\
\bottomrule
\end{tabular}%
}
\end{table}
\endgroup

\subsection{Multi-target TSE}
\label{results:multi-target}

In the multi-task TSE framework, the goal of the model is to extract one or more sources of interest from the mixture. Multi-target training involves using multi-hot vectors to denote multiple target sounds. During training, 1-3 foreground sounds are chosen randomly as targets, helping the model learn to handle multiple targets simultaneously. During evaluation, the models are applied to separate 1, 2 or 3 target sounds from the mixtures. 

Table~\ref{table_performance} presents the results of causal multi-target sound extraction where the target includes up to three sounds of interest. The results indicate that all of the proposed models, pcTCN, iCATSE, and eCATSE, consistently outperform a size- and latency-matched Waveformer model processing 16kHz-sampled audio. Due to the access to oracle context information, eCATSE shows 
substantial performance gains over all the other considered models, especially as the number of target sources to extract increases (avg. +3.55 dB SI-SNRi compared to the Waveformer). 

The proposed iCATSE model provides the best performance of all the models not relying on the oracle context. The iCATSE model consistently yields better performance across all the considered target setups and average scores, as compared to pcTCN (avg. +0.28 dB SI-SNRi) and Waveformer (avg. +0.66 dB SI-SNRi). Finally, the pcTCN model alone also outperforms the Waveformer across the board (avg. +0.38 dB SI-SNR). We speculate that some of the performance benefits may originate from the pervasive conditioning across the model, as opposed to providing the cue only once at the model input, as Waveformer does. However, it's important to note that our TCN-based architecture also differs from the hybrid DCC-Transformer employed by Waveformer.

\subsection{Single-target TSE}

Table~\ref{table_comparison} presents results for the models re-trained exclusively to perform a single-target sound extraction task. The setup is analogous to~\ref{results:multi-target}, but the models were trained and evaluated using only one target source (i.e., strictly one-hot hint vector, instead of potentially multi-hot). The evaluation data is the same as in the \textit{1 target} configuration from Table~\ref{table_performance}.

Interestingly, all the considered models processing 16 kHz audio perform slightly worse on single-target sound extraction, as compared to the multi-target training setup (Table~\ref{table_performance}). While unexpected, the slight performance gap could be explained by multi-target training acting as a regularizer, making the training more challenging and the resulting models more resilient. 

The largest performance gap in going from multi- to single-target training setup is obtained for the proposed context-aware models (i.e. iCATSE, eCATSE). While they both outperform Waveformer (iCATSE: +0.31 dB SI-SNRi, eCATSE with oracle context: +2 dB SI-SNRi), they are affected the most by reducing the number of targets in the training setup. As a result, pcTCN, with no context awareness, yielded the highest SI-SNRi, and 2nd highest SNR, amongst the models not utilizing oracle context information. This indicates that multi-target training is crucial for effectively exploiting context information either directly provided to (eCATSE), or implicitly learned by the model (iCATSE).

\section{Conclusion}

In this work, we present a novel method for context-aware, causal, low-latency TSE. Our model leverages the fact that the TSE task can be compartmentalized into target source identification and separation tasks. We confirm the benefits of the proposed context-awareness by first showing that explicit provision of the oracle context to the model substantially improves its performance (eCATSE). To relax the assumption about the oracle context, we subsequently propose the iCATSE model with multi-objective training. The training constitutes joint identification of the sound classes present in the input mixture and separation of target source(s) indicated by the cue. Using objective metrics we show that both eCATSE and iCATSE models outperform Waveformer, the current state-of-the-art low-latency, causal TSE model, thus showcasing the utility of context awareness. We highlight that multi-target TSE training is necessary to maximize the utility of context information, even if downstream inference is done using only a single target. While we used one- or multi-hot vectors to condition our models, the proposed methods can be easily adapted to work with other conditioning modalities such as text or video. Future work should also address reducing the size of the models to enable its deployment on low-resource audio streaming platforms, such as wearables. 

\bibliographystyle{IEEEtran}
\bibliography{mybib}

\end{document}